# Review on Fragment Allocation by using Clustering Technique in Distributed Database System

[1] Priyanka Dash, [2] Ranjita Rout, [3] Satya Bhusan Pratihari, [4] Sanjay Kumar Padhi

[1] Lecturer in CSE Department, GITAM, Bhubaneswar, Odisha, India

[2] Assistant Professor, CSE Department, PIET, Rourkela, Odisha, India

[3] Assistant Professor, CSE Department, INDUS College of Engineering, Bhubaneswar, Odisha, India

[4] Associate Professor, CSE Department, KIST, Bhubaneswar, Odisha, India

## Abstract

Considerable Progress has been made in the last few years in improving the performance of the distributed database systems. The development of Fragment allocation models in Distributed database is becoming difficult due to the complexity of huge number of sites and their communication considerations. Under such conditions, simulation of clustering and data allocation is adequate tools for understanding and evaluating the performance of data allocation in Distributed databases. Clustering sites and fragment allocation are key challenges in Distributed database performance, and are considered to be efficient methods that have a major role in reducing transferred and accessed data during the execution of applications. In this paper a review on Fragment allocation by using Clustering technique is given in Distributed Database System.

## 1. Introduction

Database Technology has become prevalent in most business organization. A database is a model of structures of reality.. A centralized database has all its data on one place. As it is totally different form distributed database which has data on different places. In centralized database as all the data reside on one place so problem of bottle-neck can occur, and data availability is not efficient as in distributed database. Performance degradation as number of remote sites grew, High cost to maintain large centralized DBS, Reliability problems with one, central site. A Distributed database is a database that is under the control of a central database management system (DBMS) in which storage devices are not all attached to a common CPU. It may be stored in multiple computer located in the same physical location, or may be dispersed over a network of interconnected computers. There are multiple sites (computers) in a distributed database so if one site fails then system will not be useless, because other sites can do their job because same copy of data is installed on every location. Fueled by the advances in telecommunication, distributed database systems (DDB) are becoming more affordable and useful. Ceriand pelagatti [2] defined a distributed database as A collection of data that logically belongs to the same system but is spread over the of Distributed database system (DDS) is a collection of sites connected by a communication network, in which each site is a database system in its own right, but the sites have agreed to work together, so that a user's own sits can access data anywhere in the network exactly as if the data were all stored at the user's own site [4]. Distributed databases (DDB) have been developed to meet the information needs of business organization engaged in distributed operations. Such organizations typically have facilities (sites) that have one or more computer systems (nodes) connected via some communications network (links). Users at each node have their own set of information requirements. Some of these involve data that is unique to users at a single node. Others require data is shared among users at multiple nodes.

## 2. Distributed Database Design

[1] gave a similar dentition: that a distributed database system is a collection of multiple logically interrelated databases distributed over a computer network. A distributed data base ma System (DDDMS) is then defined in [1]as the software that provides that management of the distributed database system and makes the distribution transparent to the users. [2] emphasized that the data at different its must have properties that tie them together, and that access to the files should behavior common interface. "[1] explained that the logically related files, which are individually stored at each site of a computer network, are not enough to form a distributed database. There needs to be a structure among them. They explained that physical distribution means that data does not reside at the same site in the same processor. It is pointed unit[1]that physical distribution does not necessarily imply that the computer systems are geographically distributed. The sites among the network could even have the same address. They could be in the same room, but the communication between the miss done over a Network instead of shared memory, and the communication network is the only shared resource.

## 2.1 Design Techniques: Fragmentation and allocation

The primary concern of a DDS is to design the fragmentation and allocation of the underlying data which is studied extensively in the literature [3,6]. Fragmentation and allocation are the most important elements of a distributed database design phase. They play important roles in the development of a cost efficient system[1]. To realize the benefits of a distributed database system, the first step is to partition the database into a number of on overlapping fragments and allocate these fragments to the various nodes or workstations in the DDS. Starting with the work by Chu, allocation of database using mathematical modeling techniques[7].

**Fragmentation:** A single database needs to be divided into two or more prices such that the combination of the prices yields the original database without any loss information. Each resulting piece is known as a database fragment. A fragment (horizontal, vertical) of a database object in an object-oriented database system contains subsets of its instance objects (or class extents) reflecting the way applications access the database objects. Allocating well defined fragments of classes to distributed sites has the advantage of minimizing transmission costs of data to remote sites as will as minimizing retrieval time of data needed locally. A re-fragmentation of the data is needed when application access and schema information have undergone sufficient changes. The importance of fragmentation in distributed database and subsequent allocation to distributed sites (relations or classes) has been argued by many works [5]. Most distributed database designs are static based on a priori probabilities of queries accessing database objects in addition to their frequencies which are available during the analysis stage. It is more effective for a distributed system to determine whenever re-fragmentation is necessary.

**Allocation:** Each fragment must be allocation to allocation in the distributed environment such that the system functions effectively and efficiently. Data allocation technique is used to determined the best location for the data. In the structure of distributed database system, the data should be placed inappropriate locations based on its usage requirements in order to increase local processing, Lower the data transmission among sites, and hence reduce the cost of data processing and increase the efficiency of the entire network.

Different Allocation Strategies:

i. Centralized date allocation: Entire database is stored at one site.
ii. Partitioned date allocation: Database is divided into several disjoint parts (fragments)and stores at several sites.
iii. Replicated data allocation: Copies of one or more database fragments are stored at several sites.

## 2.2 Fragment Allocation Problem

Before beginning our exploration of fragment allocation problem must be clearly defined. Here, we will only address a WAN environment since the impact of storing fragment copies on the sites of a LAN is not very significant.

Assume that we have a WAN consisting of sites $S\{S1,S2…Sm\}$, on which a set of transactions = $\{T1, T2,…,Tq\}$ is running, and a set of fragments $F=\{F1,F2,….Fn\}$, into which all global relations have been partitioned during the fragmentation phrase of distributed database design. To make the allocation problem more general, We consider that it involves not only determining the number of copies of each fragment, but also finding the optimal allocation of each fragment copy in F to S, according to the information given by the network and T. As for the definition of optimality, there are two different measures in general:

1. Minimal cost: The cost function consists of the cost of storing each Fj on site Sk, the cost of querying Fj at site Sk, the cost of updating Fj at all sites where it is stored, and the cost of data communication.
2. Performance: Two well-known strategies are to minimize the response time and to maximize the system throughput at each site.

## 2.3 Objectives of Fragment Allocation

➢ Improved Reliability and Availability

- Reliability to system live time means system is running efficiently most of time. if site fails, request can be routed to replicated date.
- Availability is the probability that the fragment is continuously available during time interval. A higher degree of availability for read only application is achieved by stored by multiple copies of the same information.

➢ Minimal Communication Cost

- Data located near user site of frequent use site of frequent use, which decrease communication cost.
- The cost function consists of the cost of storing each Fj on site Sk, the cost of querying Fj at site Sk, the cost of updating Fj at all sites where it is stored, and the cost of date communication.

➢ Improved Performance

- The performance increases when the fragment is stored at node for which they frequently accessed.

- A distributed DBMS fragments the database to keep data closer to where it is needed most. This reduces data management (access and modification) time significantly.
- Improve performance Two well-known strategies are to minimize the response time and to maximize the system throughput at each site.

Distributed database design involves the following interrelated issues: (1) how a global relation should be fragmented, (2) how many copies of a fragment should be replicated, (3) how fragments should be allocated to the sites of the communication network and [4] what the necessary information for fragmentation and allocation is [3]. Data allocation in distributed database systems is difficult as compared to allocate date as Fragmented, Replicated and Centralized [3]. Huang and Chen [4], shown that are two different measures for the definition of optimality:

1. Minimal cost: The cost function consists of the cost of storing each Fj on site Sk, the cost of querying Fj at site Sk, the cost of updating Fj at all sites where it is stored, and the cost of data communication.

2. Performance: Two well-known strategies are to minimize the response time and to minimize the system throughput at each site.

Optimality means, one should be looking for an allocation scheme that, for answers user quires in minimal time while keeping the cost of processing minimal.

2.4 Alternative Design Strategies

The design of a distributed database system involves making decisions on the architecture of DDBMS . Two major strategies proposed by Ceri and pelagatti [9] for designing distributed databases are : top-down approach and bottom-up approach. In the case of tightly integrated distributed database design proceeds stop-down form requirements analysis and logical design of the global database to physical design of each local database. In the case of distributed multi database systems, the design process is bottom-up and involves the integration of exiting databases. But real applications are rarely simple enough to fit nicely in either of these approaches. The two approaches may need to be applied together to complement each other [1].

2.4.1 Top-down Approach

In the top-down approach, the process starts with a requirement Analysis that defines the environment to the system and elicits both the data and processing needs of all potential database users [8]. The requirements analysis also specifies where the final system is expected to stand with respect to the objectives of the DDBMS. The objective is defined with respect to performance, reliability and availability, economics, and expandability (flexibility).

The requirement documents are the inputs to two parallel activities: view design and Conceptual design. The outputs of view design are the user, and the output of conceptual design is entity types and relationship types which are used to construct an externals schema.

2.4.2 Bottom-Up Approach

Ceriand pelagattin [9] and "Ozsuand Valduriez [1] stated that top-down design is suitable for the systems which are developed from scratch. But when the distributed data base is developed as the aggregation of exiting databases, it is not easy to follow the top-down approach. The bottom-up approach, which starts with individual local conceptual schemata, is more suitable for this environment [9, 1]. Ceriand Pelagatti[9] explained that the bottom-up approach is based on the integration of existing schemata into a single, global schema. Integration is the process of the merging of common data definitions and the resolution of conflicts among different representation that are given to the same data. The global conceptual schema is the product of the process [1]. Ceriand pelagatti[9]concluded that there are three requirements for bottom-up design.

1. The selection of a common database model for describing the global schema of the database.
2. The transaction of each local schema into the common data model.
3. The integration of the local schema into a common global schema.

2.4.3 The Objective of the Design of Distributed Database

Several objectives that should be taken into account in the design of distribution are presented in [2]:

- In a distributed database system one of the major costs is associated with communication. To minimize communication costs, one goal of DDBMS is to achieve processing applications locally. The degree of local processing can be maximized by distributing data, therefore minimizing transaction costs. To achieve this goal, the data should be kept as close as possible to the applications which use them. The advantage of processing applications locally is not only the reduction of remote access costs, but also increased simplicity in controlling the execution of the application.

- The availability and fault tolerance of read-only applications can be improved by storing multiple copies of the same information at different sites. When one site of the database is down or the community link for that site is broken, the

- system can still execute the applications by accessing the other copies of the information.

- Distributing workload over the sites is done in order to take advantage of the different powers of utilization of the computers at each site, and to maximize the degree of parallelism of execution of applications. But the trade off between processing locally and distributing workloads should be considered in the designing of data distribution.

- Database distribution should reflect the cost and availability of storage at each site. Even though the storage cost is not relevant when compared with the cost of input or output (I/O), central processing unit(CPU), and transmission costs of the applications, the limitation of available storage at each site should be considered.

During the design process of fragmentation and allocation, minimizing communication costs is the main objective. With the advance of current computer power, storage cost is not a big concern any more. The other two objectives, improving availability and fault tolerance and distributing workload, can be achieved when databases are fragmented and distributed properly among the network.

## 2.5 Clustering Technique

Clustering is a division of data into group of similar objects. Each groups consists of objects that are similar between themselves and dissimilar between objects of other groups. Clustering is a method of grouping sites according to a certain criterion to increase the system I/O performance. Fragment allocation technique describes the way in which the database fragments are distributed among the clusters and their respective sites in DDBs, attempts to minimize the communication costs by distributing the global database over the sites, increase availability and reliability where multiple copies of the same data are allocated, and reduces the storage overheads. Clustering as a technique to achieve high data density. Another definition of clustering is a grouping of objects together. If a use case requires objects A, B and C to operate, then those objects should be co-located for optimal data density. If upon loading the database, those objects are physically allocated close to one another, then we say we have clustered those objects. Assume that the size of the three objects combined is less than the size of a physical database page. The clustering leads to high data density because when we fetch the page with object A, we will also get objects B and C.

2.5.1 Clustering Techniques: Isolate Index

An index on a collection yields faster query performance. If the index is placed, as it is by default, in the same physical location as the collection itself and the items in the collection, then poor locality of reference the index items is likely to occur. Restated, all the items which are related to the index should be place in close proximity to one another. All items related to the collection (excluding the index) should be place in close proximity to one another. The index items should not be interspersed with the collection nor the items contained within the collection. Therefore, if we cluster the index and all the index items together in an area that is physically distinct from the collection and/or the items in the collection, the index will be faster to fetch. If there are no non-index items interspersed with the index items, then fewer pages will be required to fetch the index to perform a query or update the index. Another "side" benefit of clustering index items in an isolated area is less fragmentation. Why? If the index were interspersed with the collection and items in the collection and then the index were dropped (most likely so that it could be regenerated) the physical locale where the index items had bee would be fragmented. Fragmentation lowers data density. By definition, if you fetch a page and some space on that page is empty due to fragmentation, then you have lower data density.

2.5.2. Clustering Techniques: Object Pooling

Object pooling is a type of clustering that puts all instances of a certain type into one physical location. Imagine a scenario were an application expected to have 100 bank objects. An array, or pool, of 100 objects would be allocated at system startup. As bank objects were needed, a slot in the pool would be assigned for use by that particular bank. As banks are deleted, the slot is marked as available for reuse. This technique gives continuous storage for all objects of the type in the pool. Because deleted slots are reused, fragmentation is less likely. The lower fragmentation in combination with contiguous space leads to higher data density.

2.5.3. Clustering Techniques: Object Modeling

Object modeling is a technique that involves changes to the object model. In this category, we have four distinct methods:

- Head Body Split
- Date Member Ordering
- Collection Representations
- Virtual Keyword

**Application of Clustering**
- Data Mining
- Text Mining
- Information Retrieval
- Statistical Computational Linguistics
- Corpus-based Computational Lexicography

## 2.6 Optimal Algorithm

In distributed database systems, the performance increases when the fragments are stored at the nodes from which they are most frequently accessed. The problem is to find this particular node for each fragment. Counting the accesses of each node to a fragment offers practical solution. Having the highest access value for a particular fragment, a node could be the primary candidate to store the fragment.

2.6.1 Algorithm

**Step 1.** For each stored fragment, initialize the access counter rows to zero. ($S_{ik}=0$ were k=1,..,n).

**Step 2.** Process an access request for the stored fragment.

**Step 3.** Increase the corresponding access counter of the accessing node for the stored fragment.

**Step 4.** If the accessing node is the current owner, go to step 2.

**Step 5.** If the counter of a remote node is greater than the counter of the current owner node, transfer the ownership of the fragment together with the access counter array to remote node. (if $S_{ix}>S_{ij}$, send fragment i to node x)

**Step 6.** Go to step 2.

There are two inherent properties introduced by the optimal algorithm. First one is the *ownership property,* that is, for each fragment; the node with highest access counter value is the current owner node of the fragment, in which case the fragment is stored in this node. The second one, namely *migration property*, dictates that for any fragment the ownerships transferred to a new node, if the access counter value of the new node exceeds the access counter value of current owner node. In this case, this particular fragment migrates and is stored in this new owner node. In other words, the owner node of the fragment changes. An advantage of the optimal algorithm is the central node independence. That is, since each node runs the algorithm autonomously, there is no central node dependence. Every node is of equal importance. Whenever one node crashes, the algorithm may continue its operation without the fragments stored in the crashed node. There are two drawbacks associated with the optimal algorithm. First one is the *potential storage problem*. As the fragment size decreases and/or the number of nodes increases, the size of access counter matrix increases, which in turn results in extra storage space need for the access counter matrix. For instance, if the fragment size is one record and the number of nodes is 500, then for each record an array of 500 access counter values should be stored. In some cases, this access counter array size may exceed the record size. The second drawback is the *scaling problem* for the data type that stores the access Counter values. Since access counter values are continuously increasing, this problem may result anomalies. For example, if one byte is chosen to store the counter values, then a value greater than 255 cannot be stored in this data type.

## 2.7 Threshold Algorithm

A new algorithm namely the threshold algorithm, which overcomes the disadvantage of the optimal algorithm, is proposed for dynamic data allocation in distributed databases. The threshold algorithm reallocates data with respect to changing data access patterns. The algorithm is analyzed for a fragment using simulation. The threshold algorithm is especially suitable for a DDS where data access pattern changes dynamically. In some cases, due to extra storage space need, it could be very costly to use the optimal algorithm in its original form. For a less costly algorithm, the solution is to decrease the need for extra storage space. The heuristic threshold algorithm in this paper serves this purpose. Let the number of nodes be n and let $X_s$ denote the access probability of a node to a particular fragment. Suppose the fragment is stored in this particular node (i.e. it is the owner node). For the sake of simplicity, let $X_d$ denote the access probability of all the other nodes this particular fragment. The owner does local access, whereas the remaining nodes do remote access to the fragment. The probability that the owner node does not access the fragment is *(n-1)*$X_d$. The probability that the owner node does not perform two successive accesses is *[(n-1)*$X_d$*]* 2. Similarly, the probability that the owner node does not perform m successive accesses is [(n-1) $X_d$] m. Therefore; the probability that the owner node performs at least one access of successive accesses is 1-[(n-1) $X_d$] m.

2.7.1 Algorithm

**Step 1.** For each stored fragment, initialize the counter values to zero. (Set $S_i$ =0 for every stored fragment i).

**Step 2.** Process an access request for the stored fragment.

**Step 3.** If it is a local access, reset the counter of the corresponding fragment to 0 .Go to step 2.

**Step 4.** If it is a remote access, increase the counter of the corresponding fragment by one.

**Step 5.** If the counter of the fragment is greater than the threshold value, reset its counter to zero and transfer the fragment to the remote node. (If, $S_i>t$, set $S_i = 0$ and send the fragment to remote node)

**Step 6. Go** to step 2.

An important point in the algorithm is the choice of threshold value. This value will directly affect the mobility of the fragments. It is trivial that as the threshold value increases, the fragment will tend to stay more at a node; and as the threshold value decreases, the fragment will tend to visit more nodes. Another point in the algorithm is the distribution of the access probabilities. If the access probabilities of all nodes for a particular fragment are equal, the fragment will visit all the nodes. The same applies for two nodes when there are two highest equal access probabilities.

2.7.2 Simulation Results

In the simulation, it is assumed that there are n nodes; Xs is the access probability of the owner node; Xd is the access probability of the other nodes; Os is the probability that the fragment is in owner node and Od is the total probability that the fragment is in the other nodes. Since, Os + Od = 1, investigating only Os is sufficient. The following formula shows the relation between n, Xs and Xd.

Xs + (n-1) Xd=1

Now, let us find how a change in the access probabilities and the threshold value affect the probability that the fragment is in any mode.

2.7.2.1 Change in Access Probability

When n is held constant, Xs and Xd are inversely proportional. So, it is sufficient to investigate only the change in Xs of Os. Fig. 1 shows the behavior of Os as a function of Xs in a five-node system. Fig. 5 is drawn for three different threshold values, 0, 3 and 10.For the threshold of 0, Os is a linear function of Xs with a slope of 1. This means that when the threshold is 0, the access probability of a node directly gives the probability that the fragment is in the corresponding node.

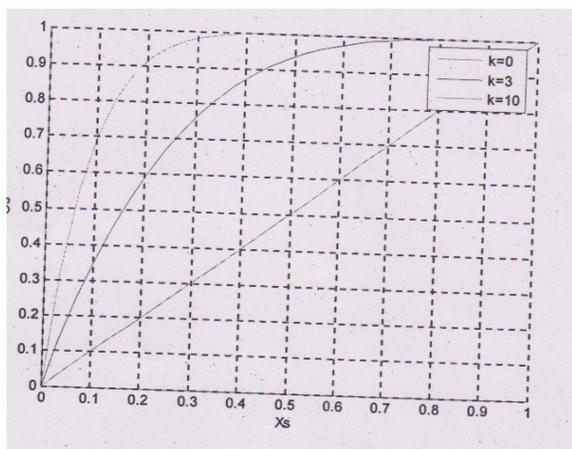

Fig.1: Change in access probability

Os as a function in a five-node system for thresholds o, 3 and 10 for threshold values of 3 and 10, notice the change in steepness of the curve.

2.7.2.2 Change in Threshold Value

Threshold can take only non-negative integer values. Fig.1 shows the behavior of Os as a function of t in a five-node system. Fig.2 is drawn for five different access probabilities Xs of 0.28, 0.24, 0.2, 0.16 and 0.12. For 0.28 and 0.24, Os converges to one. This is because Xs > Xd. Nothing the change in steepness of two curves, it converges faster for greater access probabilities.

For 0.2, Os is constant at 0.2. This is because Xs = Xd. In this case, the access probability of a node directly gives the steady-state probability that the fragment is in the corresponding node.

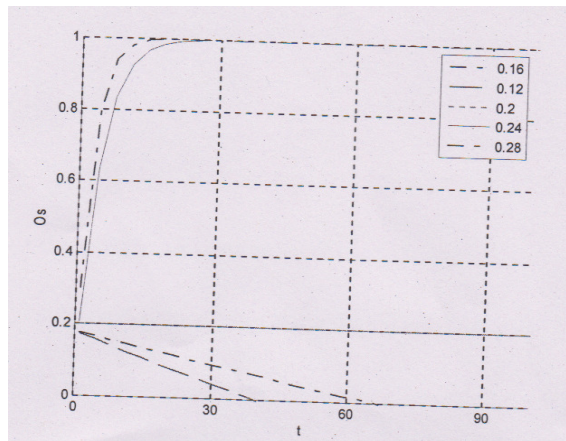

Fig.2: Change in threshold value

Os as a function of t in a five-node system for Xs values of 0.28, 0.24, 0.2, 0.16 and 0.12 .For 0.16 and 0.12, O converges to zero. This is because Xs < Xd . Noticing the change in steepness of two curves, it converges faster for smaller access probabilities. In paper [14], a new dynamic data allocation algorithm, namely threshold algorithm, for non-replicated DSSs is introduced. In the threshold algorithm, the fragments, previously distributed over a DDS, are continuously reallocates according to the changing data access patterns. The behavior of a fragment, in reaction to a change in access probabilities or to a change in threshold value, is investigated using simulation. It is shown that the fragment tends to stay at the node with higher access probability. As the access probability of the node increases, the tendency to remain at this node also increases. It is also shown that as the threshold value increases, the fragment will tend to stay more at the node with higher access probability. Threshold algorithm can be used for dynamic data allocation to enhance the performance of non-replicated DSSs. For further research, the algorithm can be extended to use on the replicated DSSs.

### 2.8 NNA Algorithm

This algorithm is very suitable for DDS in the networks which have low bandwidth and frequent requests for a data fragment come from different sites by providing data clustering. The simulation results show that for complex and large networks where the request for fragments generates more frequently or the fragment size is large, the NNA algorithm provides better response time and spends less time for moving fragments in the network. A major cost in executing queries in a distributed database system in the data transfer cost incurred in transferring relations (fragments) accessed by a query from different sites to the site where the query is initiated. The objective of a data allocation algorithm is to determine an assignment of fragments at different sites so as to minimize the total data transfer cost incurred in executing a set of queries. This is equivalent to minimizing the average query execution time, which is of primary importance in a wide class of distributed conventional as well as multimedia database systems. In this algorithm, we are going to address the problem of optimal algorithm. In NNA algorithm, the requirement for moving a fragment is obtained as in optimal algorithm. But, the destination of moved data is different. In our method we consider the network topology and routing for specifying destination. In other words the destination of the moved fragment is the neighbor of the source which is exists in the path from the source to the node with highest access pattern. Any routing algorithm can be used but we use link-state routing algorithm.

By using this approach we avoid from frequently moving data because finally the fragment will be placed in a node which has average access cost for nodes that using it. So, delay of movement will be reduced. Furthermore, the response time also will be improved. Another aspect of NNA algorithm is that the fragments which are used by a node or neighbors of a node can be clustered. By using this clustering we can effectively respond to the requests.

For example according to the Fig.3 assume that node G, H, I and E frequently send a request for a fragment i which is on the node A. According to our algorithm, after that number of requests exceeded from the predefined threshold, we move the fragment i to node C. If the requests are continued after fragment migration, we move the fragment to node B. This approach will be continued until the fragment reaches to node G. By placing the fragment on node G, the requests from G, H, I and E will be responded with less delay but not with minimum delay. In this step, data is in a stable state. After this step, if one of the nodes H, G and E request the data more frequently than the other nodes, the fragment will be placed in it. By sending the fragment to the nodes that request it with the predefined threshold, the data will be migrated frequently from one node to the other node and it takes a lot of time and also responses will be send by a lot of delay when the data is moving. Using NNA approach, avoids form these problems with trade off of providing less delay not minimum delay.

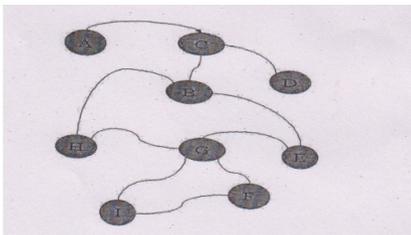

Fig 3.The topology of the experiment

## 2.9 FNA algorithm

Fuzzy Neighborhood Allocation (FNA) algorithm. This algorithm is based on NNA algorithm, but it is different with NNA in its approach for the selection of destination node for moving fragments and the recognition of oscillation conditions. It detects oscillation conditions through evaluation of differentiation of fragment access patterns. It also chooses the destination node according to summation of access pattern fuzzy vectors. The gain of the proposed fragment allocation algorithm in queries response time is greater than its execution costs. FNA algorithm will be more beneficial in situations that, distributed database system has low bandwidth or high delay links.

All of above algorithms use crisp method to move data along network paths. Estimating time and place (destination) of a data fragment depends on various parameters such as access pattern, bandwidth of network links and etc. Another condition which mentioned algorithms have ignored, is oscillation condition, which is considered as a frequent condition in traditional distributed databases. Oscillation conditions caused by alteration of fragment requests between two or more sites, take place in distributed databases such as car manufacturing companies dealership networks, ATMs and financial distributed networks and  etc. Tracking fragment migrations in distributed database system under oscillation conditions shows that fragments oscillate between sites and these excessive fragment migrations may affect system loads adversely.

Unnecessary fragment migration under oscillation condition must be avoided through an appropriate solution in order to maintain fair and reasonable performance levels. This algorithm, based on NNA[15], has different strategy in selection of nodes for data movements of migrating fragments. Proposed algorithm, enable to detect oscillation of requesting site for a specific data fragment, prevent data fragment to migrate reading to oscillation conditions triggers. The fuzzy approach detects stressed conditions via differentiation of access pattern to a special data fragment. Data fragments are to behave more stable in such oscillation circumstances and excessive data fragment migrations will be avoided.

2.9.1 Fragment Size

For small fragments the average time spent for moving data in FNA algorithm is larger than NNA and for larger fragments this is reversed. The reason is that for small fragments the cost of moving data to destination node is low and so, the movement of fragments takes more time and also increase the network traffic. So, less movement will produce some advantages that overcome the access cost. Avoidance of oscillation condition in FNA leads to have less traffic and saving in network resources such as bandwidths. In FNA destination of a data   fragment is chosen according to access pattern overall system. So, we direct our fragments more effective and this will be valuable in larger fragments.

### 2.9.2 Query Production Rate

In query production rate we neglect some of transactions in our benchmark to evaluate our algorithm in different load condition. As the rate of query is increased, the delay of response and also the average time for fragment movement is decreased. Because the high rate of query causes each fragment find its proper owner node sooner and stays on it. So, the delay of response for a fragment will be decreased. As shown, except the situation where the rate of query production is very low, FNA algorithm performs better than optimal algorithm. Deliberation in choosing destination of data fragment results avoids idle fragment movements. Less traffic is achieved by preventing oscillation condition.

### 2.9.3 Number of Active Nodes

FNA shows that as the number of active nodes(the nodes generating queries) in the network increases, the difference between FNA an NNA algorithm appears better. In our experiment we change the number of active nodes from 2 to 8. To change the number of active nodes, we neglect some of transactions according to the node, which made the transaction. According to the results we conclude that in larger networks FNA algorithm responds to a request with much lower delay than NNA algorithm. If the number of active nodes in the network increases, the average time spent for moving fragments in FNA algorithm is less than NNA algorithm. But this conclusion needs to be experimented on more complex network topologies. In our proposed algorithm, oscillations detected and recognized via a simple fuzzy inference engine. Recognizing oscillation condition leads to avoid oscillating data fragments between sites. So, idle data movement is decreased. Deliberation in fragment movement is another aspects of our proposed algorithm.

Mrs. Priyanka Dash is working as Lecturer in Computer Science and Engineering Department in GITAM, BBSR, Odisha, India.Her area of interest is Distributed Database System.

**Mrs.Ranjita Rout** is working as Assistant Professor in Computer Science and Engineering Department in PIET Rourkela, Odisha, India. Her area of interest is Distributed Database System and Image Processing.

**Mr. Satya Bhusan Pratihari** is working as Assistant Professor in Computer Science and Engineering Department in INDUS College of Engineering, Bhubaneswar, Odisha, India. His area of interest is Adhoc Network, Wireless Sensor Network and Distributed Database System.

**Mr.Sanjay Kumar Padhi** is working as Associate Professor in Computer Science and Engineering department in KIST, Bhubaneswar, Odisha, India. His area of interest is Wireless Sensor Network, Adhoc Network, Data Mining, Distributed Database System, Neural Network, Software Engineering, cloud computing etc.